\documentclass[11pt]{article}
\usepackage{amsmath}
\usepackage{amsfonts}
\usepackage{amssymb}
\usepackage{graphicx}
\usepackage{subfigure}
\usepackage{graphicx}
\usepackage{dcolumn}
\usepackage{bm}

cm \headheight=0. mm

\def\be{\begin{equation}}
 \def\ee{\end{equation}}
\def\bea{\begin{eqnarray}}
\def\eea{\end{eqnarray}}

\def\n{\nonumber}
\def\l{\label}
\def\ov{\over}

\def\R{\rho}

\begin{document}
\begin{center}
\LARGE {Quark-hadron phase transition in a chameleon Brans-Dicke model of brane gravity }
\end{center}
\begin{center}
{\bf $^{a,b} $Kh. Saaidi\footnote{ksaaidi@uok.ac.ir}},
{\bf $^{c}$A. Mohammadi\footnote{abolhassanm@gmail.com}},
{\bf$^a$T. Golanbari\footnote{teagol@gmail.com}},
{\bf$^a$H. Sheikhahmadi\footnote{h.sh.ahmadi@uok.ac.ir}}
{\bf$^b$B. Ratra\footnote{ratra@phys.ksu.edu}}
\\
{\it$^a$Department of Physics, Faculty of Science, University of
Kurdistan,  Sanandaj, Iran}\\
{ \it $^b$Department of Physics, Kansas State University,116 Cardwell Hall, Manhattan, KS 66506, USA.}\\
{\it $^c$Young Researcher Club, Larestan Branch, Islamic Azad University, Lar, Fars
Province, Iran}\\

\end{center}
 \vskip 1cm
\begin{center}
{\bf{Abstract}}
\end{center}
 \ \ \ In this work, the quark-hadron phase transition in a chameleon Brans-Dicke model of brane world
cosmology within an effective model of QCD is investigated. Whereas, in the chameleon Brans-Dicke
model of brane world cosmology, the Friedmann equation and conservation of density energy are
modified, resulting in an increased expansion in the early Universe. These have important effects on
quark-hadron phase transitions. We investigate the evolution of the physical quantities relevant to
quantitative descriptions of the early times, namely, the energy density, $\rho$, temperature, $T$, and the scale
factor, $a$, before, during, and after the phase transition. We do this for smooth crossover formalism in
which lattice QCD data is used for obtaining the matter equation of state and first order phase transition
formalism. Our analyses show that the quark-hadron phase transition has occurred at approximately one
nanosecond after the big bang and the general behavior of temperature is similar in both of two
approaches.\\



\section{Introduction}

As the  universe expanded and cooled it passed through  a series of symmetry-breaking phase transitions which can generate  topological defects. Here we study the quark-gluon (QG)  to hadron phase transition.  This early universe  phase transition  has been  studied in detail  for over three decades   \cite{18, 19, 20, 21, 22, 23}.
It could be a first, second, or higher order  phase transition. In addition, the possibility of no phase transition was considered  in Ref. \cite{31}. The order of the phase transition  depends strongly  on the mass and flavor of the quarks.

For an early study of a first-order quark-hadron phase transition in the expanding universe see Ref. \cite{32}. As the  deconfined quark-gluon plasma cools below the critical temperature $T_c , = 150$ MeV, it becomes energetically favorable to form color-confined hadrons (mainly pions and a few of neutrons and protons, due to the conserved net baryon number). However, this new phase does not form immediately. As is characteristic of a
first-order phase transition, some supercooling is needed to overcome the energy expense of forming the surface of the bubble and the new hadron phase. When a hadron bubble is nucleated, latent heat is released, and a spherical shock wave expands into the surrounding supercooled quark-gluon plasma. This reheats the plasma to the critical temperature, preventing further nucleation in  regions passed through by  one or more shock fronts. Generally, bubble growth is described by deflagrations, with a shock front preceding the actual transition front. The nucleation stops when the  universe has reheated to $T_c$. After that, the hadron bubbles grow at the expense of the quark phase and eventually percolate or coalesce. The transition ends when all quark-gluon plasma has been converted to hadrons, neglecting possible quark nugget production. The physics of the quark-hadron phase transition, as well as the cosmological implications of this process, has been extensively discussed in the framework of general relativistic cosmology in Refs. \cite{33, 36, 37, 40, 41, 42, 43}.

As an alternative to general relativity,  the scalar-tensor theory was conceived originally by Jordan,
who  embedded a four-dimensional curved manifold in five-dimensional flat space-time \cite{1}. Scalar-tensor models
include  a  scalar field, $\phi$, with non-minimal coupling
to  the geometry in the gravitational action, as introduced  by
Brans and Dicke (BD) \cite{2}. Brans-Dicke models
 have  proved to be useful as a setting for discussing  some of the outstanding
puzzles in cosmology {\cite{3,4}}.
The  mechanism that creates a  non-minimal scalar field
coupling to the geometry could  also lead to a
coupling between the scalar and matter fields. Two such examples  are a generalization  of  quintessence \cite{23a}   and the chameleon  mechanism
\cite{7, 6}. The scalar field in the generalized  quintessence scenario has a very small  mass and couples to matter  with gravitational
strength. The authors of Ref. \cite{7}  study   a chameleon mechanism where the scalar
field couples directly   to matter with order unity strength. In this mechanism  the mass of the scalar field
depends on the local mass density. The chameleon
proposal provides  a way to generate  an effective mass for  a light scalar
field via   field self-interaction and  the interaction between  matter
 and scalar fields.  When the chameleon coupling is used in  the Brans-Dicke model, this is  called the
chameleon-Brans-Dicke model \cite{8}.
Solar system   observational  constraints on  the  chameleon-Brans-Dicke model have been   studied  in Ref. \cite{12}.

Over the past decade the possibility that our four dimensional  universe is a brane  embedded in a higher
dimensional space-time has attracted considerable interest \cite{13}. This  scenario has been investigated for the case in  which the bulk is  { five} dimensional  and it has been  shown that it  can result in  a theory of gravity which mimics purely four-dimensional gravity, both with respect to the classical gravitational potential and with respect to gravitational radiation \cite{14}.

 Of interest in the present
study are  brane-world models in the context of chameleon-Brans-Dicke (CBD) gravity. Interestingly, it will
be show that in such a scenario, and in the presence of a CBD field in the bulk, due to non-minimal coupling between  the scalar field and matter, the energy conservation equation on the brane  for  matter fields is modified. It is   of interest to study the   quark-gluon    to   hadrons  phase  transition  in the context of the  CBD brane world theory of gravity. The  quark-hadron
phase transition in the context of conventional brane-world gravity and in  Brans-Dicke  brane-world gravity have  been studied in Refs. \cite{15, 16, 17}.

 Recently, based on the particle physics motivation, there has been interested in the possibility of  energy exchange between the brane and
  bulk. Observational constraints on
cosmological models in the brane-world scenario in which the bulk is not empty, and that allow for the exchange of mass-energy between the bulk and the brane have been  studied  \cite{ 141}. The  evolution of matter fields on the brane is modified due to new terms in the energy momentum tensor that describe this exchange. This model  can account for the observed suppression of the cosmic microwave background (CMB) power spectrum at low multipoles, and  in this model the observed recent cosmic
  acceleration is
attributable to the flow of matter from the bulk to the brane. The cosmological evolution of
a brane with  chameleon scalar field  and general matter content in the bulk was considered in Ref.\;\cite{142}. Also the reheating the universe in brane-world model of  cosmology with bulk-brane energy transfer has been studied in Refs.\;\cite{143, 144}.

In fact existence  an energy dissipation from the bulk scalar field into the matter field on the brane, shows  an interaction between matter and scalar field. {  But  when a chameleon scalar field interacts with perfect fluid,  this interaction  produce a fifth
force on the matter which may violate  the  equivalence principle (EP) and  creates a non-geodesic motion.
This kind of interactions have attracted much attention \cite{7, 8, 12, 142, Faraj, kh3, khaled2}.  As was mentioned earlier,
the mass of chameleon scalar field is a function of local density and  in the high density regions\footnote{Where observation and experiments are performed such as Earth.} the fifth force effects are confined to an undetectable small distances. Therefore the violation of EP is not observed in the laboratory \cite{7,kh3}.  Moreover, In \cite{khaled2} has been shown that  the motion of perfect fluid in this model is the same as the motion of perfect fluid in Einsteinian theory. In fact it is shown that for $L_m=-p$ as a Lagrangian density of perfect fluid,  the motion of perfect fluid is geodesic. Therefore we consider this model due to the following reasons.
 \begin{itemize}
 \item  The brane world theory is an outstanding motivation in cosmology.
 \item  This model can create a bulk-brane energy transfer which has been studied in early time often.
\item  The motion of perfect fluid in this model is geodesic.
\item Study of quark-hadron phase transition in this model shows some interesting results.
\end{itemize}
}
This paper is organized as follows. In Sec.\;2, we  introduce the model and derive the equations of motion. We review the first-order phase transition and consider it in  our model in Sec.\;3. In Sec.\;4  we investigate our model for a typical example.  We study the smooth crossover  approach in Sec.\;5 and Sec.\;6  summarizes   our results.


\section{General framework}
We consider the five-dimensional   chameleon-Brans-Dicke  model with action
\begin{equation}\l{1}
S=-\frac{1}{2\kappa_{(5)}^2}\int d^5x  \sqrt{-g}\Big[\phi R  -
\frac{\omega}{\phi} \partial _{A}\phi \partial ^{A} \phi - V(\phi) \Big] + \int
d^5x\sqrt{-g}  f(\phi)L_m .
\end{equation}
Here $g$ is the  determinant of the five-dimensional metric $g_{AB}$, $R$ is the Ricci scalar
constructed from  $g_{AB}$, $\phi$ is  the CBD  scalar
field, $\omega$ is a dimensionless coupling constant which determines the
coupling between gravity and $\phi$, $L_m$ is  the   Lagrangian density for the matter
fields, and
 $V(\phi)$ is the scalar field  potential energy density. Latin indices label  five-dimensional components ($A,B$ = 1, . . . , 5) and for convenience we shall set  $\kappa^2_{(5)}=8\pi G_{(5)} = 1$, where $G_{(5)}$ is the five-dimensional Newtonian gravitational constant.
 The last term
on the right hand side of Eq. (1),
$f(\phi){L}_m$, indicates non-minimal coupling between the
scalar field and matter, where $f(\phi)$ is an analytical  function
of $\phi$.

 One can obtain the gravitational and scalar  field equations of motion  by
varying  the action (1) with respect to  $g_{AB}$  and $\phi$. The gravitational field equation is
\begin{equation}\l{2}
^5G_{AB}\equiv  {^5}R_{AB} - \frac{1}{2}g_{AB}R=\frac{1}{\phi}\left( T^{\phi}_{AB}+f(\phi)T_{AB}\right),
\end{equation}
where $^5G_{AB}$ is the five-dimensional Einstein tensor, $^5R_{AB}$ is the five-dimensional  Ricci tensor.  The five-dimensional
energy-momentum tensor of the matter, $T_{AB}$,  is  given by
\begin{equation}\l{2'}
T_{AB} = {2\over \sqrt{-g}}{\delta(\sqrt{-g}L_m) \over \delta g^{AB} },
\end{equation}
 and the scalar field energy-momentum tensor, $T^{\phi}_{AB}$, is
\begin{equation}\l{3}
T^{\phi}_{AB}=\frac{\omega}{\phi} \Big[ \nabla _A \phi \nabla _B \phi -\frac{1}{2} g_{AB}\nabla _C \phi \nabla ^C \phi \Big]
+ \Big[\nabla _A \nabla _B \phi - g_{AB}\nabla _A \nabla ^A  \phi \Big] - \frac{1}{2}g_{AB}V(\phi).
\end{equation}
The scalar field equation of motion  is
\begin{equation}\l{4A}
(3\omega+4)\nabla _A \nabla ^A  \phi =  \left[ Tf(\phi)+3\tilde{f}(\phi)\phi L_m \right] +  \frac{3}{2}\phi \tilde{V}(\phi) - \frac{5}{2}V(\phi),
\end{equation}
where $T$ is the trace of $T_{AB}$ and   $\tilde{X}:=\frac{dX}{d\phi}$. Setting  $f(\phi)=1$ and $V(\phi) =0$, the above equations reduce to those  of Ref.\;\cite{17}.

{ Note that to solve  Eq.\;(\ref{4A}) we need an explicit form of the perfect fluid Lagrangian density. In \cite{khaled1} have been shown that, for perfect fluid that does not couple explicitly
to the other components of the system, there are different Lagrangian densities which are perfectly equivalent.
In fact, they have shown that, by using Eq.\;(\ref{2'}), the two Lagrangian densities $L_{m_1} = -p $ and $L_{m_2} = \rho$
give the same stress-energy tensor, moreover  the equation of
motions for all components of the system for these two different Lagrangian densities are similar. But according to \cite{khaled2}, when perfect fluid couple explicitly to the scalar field, these two perfect Lagrangian densities are not equivalent and for $L_m =-p$  the motion of perfect fluid is geodesic. Therefore in this work we choose $L_m=-p$.  }

  We consider a five-dimensional flat metric of the  form
\begin{equation}\l{5A}
ds^2=-n^2(t,y)dt^2+a^2(t,y)\gamma_{ij}dx^idx^j+b^2(t,y)dy^2,
\end{equation}
where $i, j$ = 1, 2, 3. We also assume an orbifold symmetry along the fifth direction $y=-y$.

 One can   define  the energy-momentum tensor as
\begin{equation}\l{6}
T^A_{\phantom{A} B}=T^A_{\phantom{A} B}\mid _{bu}+T^A_{\phantom{A} B}\mid _{br},
\end{equation}
where the subscripts $br$ and $bu$ refer to the corresponding energy-momentum tensors in the  brane and bulk respectively.  We assume the brane has tension $\lambda$ and  is filled with perfect fluid  matter and the bulk has no   ordinary matter. I. e.,  the  matter energy-momentum tensors   are taken to be
 \begin{eqnarray}\l{7''}
T^A_{\phantom{A} B}\mid _{br}&=&\frac{\delta(y)}{b}{\rm diag}(-\rho,p,p,p,0),\\
T^A_{\phantom{A} B}\mid _{bu}&=&{\rm diag}(0,0,0,0,0),
\end{eqnarray}
where $\delta(y)$ is the Dirac delta function. It is assumed  that the brane is held at $y$ = 0, and
\begin{eqnarray}\l{7}
\rho &=&\rho_m+\lambda,\\
p&=&p_m-\lambda,\l{7'}
\end{eqnarray}
 where the subscript $m$ denotes   matter. There are several  suggestions for  the appropriate numerical   value of the brane tension $\lambda$. From the success of  big bang nucleosynthesis $ \lambda \geq 1$ MeV$^4 $ \cite{28s}. A
much stronger bound for $\lambda$ comes from   the null results of submillimeter tests of  Newton's  inverse-square law of gravity, giving
$\lambda \geq 10^8 {\rm GeV}^4$ \cite{29s}. An astrophysical  limit on $\lambda$,   independent of  Newton's law of gravity, and cosmological limits, have  been studied in Ref. \cite{28s}, leading to   $\lambda > 5\times10^8 $ MeV$^4 $.

We assume that the five dimensional  metric (\ref{5A}) is
continuous, but the first derivative with respect to $y$ is
discontinuous, so that  the second derivative with respect to $y$
includes a Dirac delta function. Making use  of Eq. (\ref{5A}),
one can obtain the non-vanishing components of the Einstein tensor. The (0,0) component of the  Einstein equation is
\begin{eqnarray}\l{8}
^5G_{00}&=&3 \Bigg[ \frac{\dot{a}}{a} \left(
\frac{\dot{a}}{a}+\frac{\dot{b}}{b} \right) - \frac{n^2}{b^2}
\left\{ \frac{a''}{a} + \frac{a'}{a} \left( \frac{a'}{a} - \frac{b'}{b} \right) \right\} \Bigg]=\frac{1}{\phi}\left[ T^{\phi}_{00}+f(\phi)T_{00}\right],
\end{eqnarray}\l{9}
 where
  \begin{eqnarray}\l{9}
T_{\rm 00}^{\phi}&=&-\dot{\phi}\left(3\frac{\dot{a}}{a} +
\frac{\dot{b}}{b} -\frac{\omega}{2} \frac{\dot{\phi}}{\phi}\right) +
\left( \frac{n}{b} \right)^2 \left[ \phi^{\prime\prime} +
\phi^{\prime} \left( 3 \frac{a^{\prime}}{a} - \frac{b^{\prime}}{b} +
\frac{\omega}{2} \frac{\phi^{\prime}}{\phi} \right)\right]+ \frac{n^2}{2}V(\phi).
\end{eqnarray}
The $(i, j)$ component of the  Einstein equation is
\begin{eqnarray}\l{10}
^5G_{ij}&=&\frac{a^2}{b^2}\gamma_{ij}\Bigg[ \frac{a'}{a} \left(
\frac{a'}{a}+2\frac{n'}{n} \right) - \frac{b'}{b}
 \left( \frac{n'}{n}+2\frac{a'}{a} \right) +2\frac{a''}{a}+\frac{n''}{n} \Bigg] {}  \\
 & & + \frac{a^2}{n^2}\gamma_{ij} \Bigg[ \frac{\dot{a}}{a} \left( -\frac{\dot{a}}{a}+2\frac{\dot{n}}{n} \right)
 - 2\frac{\ddot{a}}{a} + \frac{\dot{b}}{b}\left( -2\frac{\dot{a}}{a}+\frac{\dot{n}}{n} \right)
- \frac{\ddot{b}}{b} \Bigg]=\frac{1}{\phi}\left[ T^{\phi}_{ij}+f(\phi)T_{ij}\right],\nonumber
 \end{eqnarray}
 where
 \begin{eqnarray}\l{11}
 T_{ij}^{\phi}=&&{\Biggr [}\Big(\frac{a}{n}\Big)^2 \Bigg\{ \ddot{\phi} + \dot{\phi} \Big( \frac{\omega}{2} { \dot{\phi} \over \phi} + 2{ \dot{a} \over a} - { \dot{n} \over n} + { \dot{b} \over b} \Big) \Bigg\}\nonumber\\
  &&- \Big(\frac{a}{b}\Big)^2 \Bigg\{ \phi'' + \phi' \Big( \frac{\omega}{2} { \phi' \over \phi} + 2{ a' \over a} - { n' \over n} + { b' \over b} \Big) \Bigg\}+ {1 \over 2} a^2 V(\phi) {\Biggl ]}\delta_{ij}.
 \end{eqnarray}
The (0,5) and (5,5) components of  the Einstein equation  are
 \begin{eqnarray}\l{12}
^5G_{05}&=&3\left( \frac{n'}{n}\frac{\dot{a}}{a} + \frac{a'}{a}\frac{\dot{b}}{b} - \frac{\dot{a'}}{a} \right)=\frac{1}{\phi} T^{\phi}_{05},\\
^5G_{55}&=&3{b^2 \over n^2}\Bigg[ -\frac{a'}{a}\left( \frac{a'}{a}+\frac{n'}{n}
\right) + \frac{b^2}{n^2} \left\{ \frac{\dot{a}}{a} \left(
\frac{\dot{a}}{a}-\frac{\dot{n}}{n} \right) \right\} \Bigg]=\frac{1}{\phi}\left[ T^{\phi}_{55}+f(\phi)T_{55}\right],\l{13}
\end{eqnarray}
where
\begin{eqnarray}\l{14}
T_{05}^{\phi}&=&\dot{\phi}^{\prime} - \dot{\phi}\left(
  \frac{n^{\prime}}{n} - \omega\frac{\phi^{\prime}}{\phi}
\right) - \frac{\dot{b}}{b} \phi^{\prime},\\
T_{55}^{\phi}&=&\ddot{\phi} +\dot{\phi} \left( 3 \frac{\dot{a}}{a} -
\frac{\dot{n}}{n} + \frac{\omega}{2}
 \frac{\dot{\phi}}{\phi} \right) - \left( \frac{n}{b}\right)^2
\phi^{\prime} \left( 3\frac{a^{\prime}}{a}  + \frac{n^{\prime}}{n} -
\frac{\omega}{2} \frac{\phi^{\prime}}{\phi} \right).\l{15}
\end{eqnarray}
The  equation of motion of the CBD scalar field is
\begin{eqnarray}\l{16}
\ddot{\phi}+\dot{\phi} \left( 3\frac{\dot{a}}{a}+\frac{\dot{b}}{b}-\frac{\dot{n}}{n} \right)&-& \left(\frac{n}{b}\right)^2 \left[ \phi '' + \phi ' \left( \frac{n'}{n} + 3 \frac{a'}{a} - \frac{b'}{b} \right) \right]= \\ &-&  \frac{n^2 }{(3\omega+4)} \left[ \left\{T f(\phi) -3p \tilde{f}(\phi) \phi \right\}  +
  \frac{3}{2}\phi \tilde{V}(\phi) - \frac{5}{2}V(\phi)\right]. \nonumber
\end{eqnarray}
In these equations, $\dot{X}:=\frac{dX}{dt}$ and $X':=\frac{dX}{dy}$.

 Since the second
derivative of the  metric includes a Dirac delta function, according to  Ref. \cite{21s}, one can define
\begin{equation}\l{17}
W''=\hat{W}'' + [W']\delta{(y)}
\end{equation}
where $\hat{W}''$ is the
non-distributional part of the double derivative of $W(t,y)$, and
$[W']$ is the jump in the first derivative across $y=0$, which is
defined by
$$[W']=W'(0^+)-W'(0^-).$$
The junction relations  can be obtain
by matching the  coefficient of the  Dirac delta function on both sides of the Einstein equation.
From the $(0,0)$ and $(i,j)$ components of the  field equation we have,
respectively
\begin{eqnarray}\l{18}
\frac{[a'_0]}{a_0b_0}&=&\frac{-1}{(3\omega+4)\phi_0} \Bigg\{ \big [p+(\omega+1)\rho\big]f(\phi_0)-p\phi_0 \tilde{f}(\phi_0) \Bigg\},\\
\frac{[n'_0]}{n_0b_0}&=&\frac{1}{(3\omega+4)\phi_0} \Bigg\{ \Big[ 3(\omega+1)p+(2\omega+3)\rho \Big] f(\phi_0)+p{\phi_0 \tilde{f}(\phi_0)} \Bigg\},\l{19}\\
\frac{[\phi '_0]}{\phi_0b_0}&=&\frac{1}{(3\omega+4)\phi_0} \Bigg\{(3p-\rho) f(\phi_0)-3p\phi_0 \tilde{f}(\phi_0) \Bigg\}.\l{20}
\end{eqnarray}
 For $f(\phi) =1 $ these equations reduce to the junction
relations of Refs. \cite{17, 30s}.

 Using the ($0, 0$) component of the Einstein field equation  for a
brane  which is located at $y=0$ and the equations which  represent the jump
conditions, (\ref{18}), (\ref{19}) and  (\ref{20}),
one can derive  the Friedmann equation
\begin{eqnarray}\l{21}
H^2&+&\Upsilon \Big( H-\frac{\omega}{6}\Upsilon \Big)\nonumber \\&=&\frac{1}{24(3\omega+4)^2 \phi^2_0} \Bigg[ \Big\{ {\omega}(3p-\rho)^2+6(2+3\omega+\omega^2)\rho^2-6\omega\rho p-12p^2\Big\}f^2(\phi_0) \nonumber \\
 &&\;\;\;\;\;\;\;\;\;\;\;\;\;\;\;\;\;\;\;\;\;\;\;\;\ +  3({3\omega}-4)p^2{\phi^2_0 \tilde{f}^2(\phi_0)}+4(3\omega +4)^2\phi_0 V(\phi_0)  \nonumber \\
 & &\;\;\;\;\;\;\;\;\;\;\;\;\;\;\;\;\;\;\;\;\;\;\;\;\;\;\;\;\;\;\;\;\;\;\;\;\;\;\;\; +6\big [(4- 3{\omega})p+2 \omega\rho)\big]p\phi_0 f(\phi_0) \tilde{f}(\phi_0)  \Bigg],
\end{eqnarray}
where $H=\dot{a}/a$ is the Hubble parameter, $\Upsilon = {\dot{\phi}}/{\phi}$, and the subscript $0$ indicates  the quantity is on the brane.

 From the  $(0,5)$ component of the field
equation,  using Eqs.\;(\ref{18}), (\ref{19}) and  (\ref{20}), we obtain  the energy conservation  equation on the brane
\begin{eqnarray}\l{22}
\dot{\rho}+3H(\rho+p)=-(p+\rho)\frac{\tilde{f}(\phi_0)}{f(\phi_0)}\dot{\phi}_0.
\end{eqnarray}
As expected, due to the interaction between the matter
and scalar field,
 the energy conservation relation is  modified. Note that in this Section we have used   $n_0=1$ and $b_0=1$, without  loss of generality.
 Using Eqs.\;(\ref{17}) and (\ref{23}) we can obtain the equation of motion for $\phi $ on the brane
\begin{eqnarray}\l{23}
{\ddot{\phi}_0}+3H\dot{\phi}_0 &=&-\frac{1}{8(3\omega+4)^2 \phi_0} \Bigg[ {4(3\omega+4)\phi_0 }\Big\{ 3\phi_0 V'(\phi_0)-5V(\phi_0)\Big\} \\
 &&\;\;\;\;
 -(3p-\rho)^2\omega f^2(\phi_0)+12p^2\phi_0^2 \tilde{f}^2(\phi_0)
-(4-3\omega)p(3p-\rho)\phi_0 f(\phi_0)\tilde{f}(\phi_0) \Bigg],\nonumber
\end{eqnarray}
where we have assumed $\hat{\phi''} =0 $.

 For simplicity,  we assume
\begin{equation}\l{28}
\phi_0 = Na_0^{n},
\end{equation}
where $N$ and $n$ are constants.  For small $n$  this ansatz  has been shown to lead to consistent results \cite{366}. With this choice
 the energy  conservation equation becomes
\begin{equation}\l{29}
\dot{\rho}+(3+n\phi_0\bar{F})H( \rho +p) =0,
\end{equation}
where $\bar{F} = \tilde{f}/f$.


\section{Quark-hadron phase transition}
 In this Section we  study a first-order quark-hadron  phase transition  in the early universe within
the  CBD brane-world scenario. For a review of a first-order  quark-hadron phase transition  see Ref. \cite{15} and references therein.

The energy density and pressure  of matter in the quark-gluon phase at temperature $T$  are \cite{15}
\begin{equation}\l{24}
\rho _q=3a_q T^4 +U(T), \ \ \ \ \ p_q =a_q T^4 -U(T).
\end{equation}
Here the subscript $q$ denotes quark-gluon matter and   $a_q=61.75(\pi^2/90)$. The potential energy density, $U(T)$,  is  \cite{21s}
\begin{equation}\l{25}
U(T)=B+\gamma_T T^2 - \alpha_T T^4,
\end{equation}
where $B$ is the bag pressure   constant, $\alpha_T = 7\pi^2/20$, and
$\gamma_T = m^2_s/4$, where $m_s$,   the mass of the strange quark,  is in the range
$m_s \in (60 - 200)$ MeV. This form of  $U$ is for a  model in which the quark fields
interact with a chiral field formed from the $\pi$ meson field and a scalar field \cite{40}.
Results obtained in low energy hadron spectroscopy, heavy ion collisions, and  from phenomenological fits of light hadron
properties, give $B^{1/4}$ between 100 and 200 MeV.

In the hadron phase one takes the cosmological fluid to be   an ideal gas of massless pions and  nucleons
described by the Maxwell-Boltzmann  distribution function, with energy density $\rho_h$ and pressure $p_h$. Hence the equation of state in the hadron phase is
\begin{equation}\l{26}
 p_h ={1 \over 3}\rho _h=a_{\pi} T^4,
\end{equation}
where $a_{\pi} =17.25 (\pi^2/90)$.

The critical temperature $T_c$ is defined by the condition $p_q (T_c) = p_h (T_c)$ \cite{31},  and,  for $m_s = B^{1/4} = 200$ MeV,  is given by
\begin{equation}\l{27}
 T_c={{\Bigg[}{\frac{\gamma_T + \sqrt{\gamma^2_T+ 4B(a_q+\alpha_T-a_{\pi})}}{2(a_q+\alpha_T -a_{\pi})}}{\Bigg]}}^{1\over 2}\approx 125   {\rm MeV}.
\end{equation}
 Since the phase transition is  first order, all physical quantities, such as  the energy density, pressure, and entropy, exhibit discontinuities across the critical curve.


\subsection{Evolution of temperature   in quark-gluon  phase (QGP)  for general $U(T)$}

In this Subsection we  study the quark-hadron phase transition  in the chameleon Brans-Dicke brane-world scenario for the potential energy density of Eq.\;(\ref{25}). The quantities   we want to trace through the quark-hadron phase transition are
the temperature $T$ and the scale factor $a$. To accomplish this  we use the equations  obtained in Sec.\;2.
In the quark-gluon phase with $T>T_c$,  from     Eqs.\;(\ref{24}), (\ref{25}) and (\ref{29}),  we have
\begin{equation}\l{30}
H= -\frac{ 2(3a_q-\alpha_T)T^2 +\gamma_T  }{2(3+n\phi_0\bar{F})a_q T^3}\dot{T}.
\end{equation}
This equation can be used to determined the scale factor as a function of $T$.
Using Eqs.\;(\ref{7''}), (\ref{7'}) and (\ref{28}), the Friedmann equation, Eq.\;(\ref{21}), becomes
\begin{eqnarray}\l{32}
H^2&=& \frac{1}{4(3\omega+4)^2\theta^2\phi_0^2} \Bigg[ 6\Big\{ (4- 3 \omega)(p_q-\lambda)+2\omega (\rho_q+\lambda) \Big\}(p_q-\lambda) \phi_0 f(\phi_0)\tilde{f}(\phi_0) \n\\
 && +\Big\{ {\omega}H_1^2 + 6(2+3\omega+\omega^2)(\rho_q+\lambda)^2 - 6\omega(\rho_q+\lambda)(p_q-\lambda)
  -12(p_q-\lambda)^2 \Big\}f^2(\phi_0)\nonumber \\
   &&\;\;\;\;\;\;\;\;\;\; +   3\Big({3\omega}-4\Big)(p_q-\lambda)^2 {\phi_0^2 \tilde{f}^2(\phi_0)}+  4(3\omega+4)^2 \phi_0 V(\phi_0)\Bigg],
\end{eqnarray}
where
$$\theta^2 = (6+6n-{\omega}n^2),$$
$$H_1 = 3p_q-\rho_q - 4\lambda.$$

Combining these equations,  one  obtains   the  expression governing the evolution of temperature  in the quark phase
\begin{eqnarray}\l{33}
\dot{T}&=&-\frac{a_q(3+n\phi_0F)T^3}{(3\omega+4)\theta\phi_0\Big[ 2(3a_q-\alpha_T)T^2 +\gamma_T \Big]}  \\
 &&\;\;\times  \Bigg[ \Big\{{\omega}T_1^2 + 6(2+3\omega+\omega^2)T_3^2
  -6\omega T_3T_2
  - 12T_2^2 \Big\}f^2(\phi_0)+ 4(3\omega+4)^2\phi_0 V(\phi_0) \nonumber \\
 &&\;\;\;\;\;\;\; + 6\Big\{ \big(4-3\omega\big)T_2 +2{\omega}T_3 \Big\}    T_2\phi_0 f(\phi_0)\tilde{f}(\phi_0)
  +3\Big({3\omega}-4\Big)T^2_2\phi_0^2\tilde{f}^2(\phi_0)  \Bigg]^{1/2}, \nonumber
\end{eqnarray}
where
$$T_1=4\alpha_T T^4 - 4\gamma_T T^2 - 4B - 4\lambda,$$
$$T_2 = (a_q+\alpha_T)T^4 - \gamma_T T^2 -B-\lambda ,$$
$$T_3=(3a_q-\alpha_T)T^4 + \gamma_T T^2 +B+\lambda .$$


\subsection{Evolution of temperature  in QGP for $U (T) = B$}

When dealing with quark confinement, one popular model is that of an elastic bag which allows the quarks to move  around freely, and the potential energy density is constant. In this case   the   equation of state  for  quark matter is $p_q = {(\rho_q -4B) / 3}.$ In this  Subsection we assume an   equation of state  of  quark matter  given by this bag model. For this case, the expression (\ref{30}) becomes
\begin{equation}\l{34}
H=- \frac{ 3\dot{T}}{(3+n\phi_0\bar{F})T}.
\end{equation}
From this relation we can obtain  an expression for the scale factor as a function of   $T$.
Also, Eq.\;(\ref{33}) reduces to
\begin{eqnarray}\l{35}
\dot{T}&=&-\frac{(3+n\bar{F}(\phi_0))T}{3(3\omega+4)\theta\phi_0  }  \\
 &&\;\;\times  \Bigg[ \Big\{{\omega}T_{10}^2 + 6(2+3\omega+\omega^2)T_{30}^2
  -6\omega T_{30}T_{20}
  - 12T_{20}^2 \Big\}f^2(\phi_0)+ 4(3\omega+4)^2\phi_0 V(\phi_0) \nonumber \\
 &&\;\;\;\;\;\;\; + 6\Big\{ 2{\omega}T_{30} +( 4 -3\omega)T_{20} \Big\}
    T_{10}\phi_0 f(\phi_0)\tilde{f}(\phi_0)
  +3\Big({3\omega}-4\Big)T^2_{20}\phi_0^2\tilde{f}^2(\phi_0)  \Bigg]^{1/2}, \nonumber
\end{eqnarray}
where
$$T_{10}=-4( B +\lambda),$$
$$T_{20} = a_qT^4 -B-\lambda ,$$
$$T_{30}= 3a_qT^4  +B+\lambda .$$
In Eq.\;(\ref{35})  the scalar field $\phi_0$ is a function of the scale factor $a$ whose  dependence on the  temperature $T$  is determined from Eq.\;(\ref{34}).


\subsection{Evolution of hadron  volume fraction   }
During the quark-hadron  phase transition $\rho_q(t)$ decreases from $\rho_q(T_c)=\rho_Q$ to $\rho_h(T_c)=\rho_H$,  but the   temperature and pressure stay constant. At the  phase transition temperature $T_c = 125$ MeV we have
$\rho_q \approx 5 \times 10^9$MeV$^4$, $\rho_h\approx 1.38 \times 10^9$ MeV$^4$,  and $p_c \approx 4.6 \times 10^8$MeV$^4$  is constant during the phase transition. Following Refs.\;\cite{15, 17, 31}, one can replace $\rho(t)$  by $h(t)$, the volume fraction of matter in the hadron phase, by
defining
\begin{equation}\l{35'}
\rho_q(t)= \Big[1+mh(\tau)\Big]\rho_Q,
\end{equation}
where $m= {\rho_H}/{\rho_Q} -1 = {\rm constant} $. At the beginning of the phase transition  $\rho(\tau_c)=\rho_Q$ and $h(\tau_c) = 0$, where $\tau_c$ is the time at the beginning of the phase transition, while at the end of the transition $\rho(\tau_h)=\rho_H$ and $h (\tau_h) = 1$,
where $\tau_h$ is the time at the end of the   phase transition.

 For $\tau > \tau_h$ the universe is in the hadronic phase. Then, from   Eqs.\;(\ref{29}) and (\ref{35'}),  we arrive at
\begin{equation}\l{36}
H=- \frac{r\dot{h}}{(3+n\phi_0\bar{F})(1+rh)}
\end{equation}
where
\begin{equation}\l{37}
r=\frac{\rho_H-\rho_Q}{p_c+\rho_Q} = {\rm constant},
\end{equation}
and
\be\l{37''}
A_0=\frac{3p_c+\rho_Q-2\lambda}{p_c+\rho_Q}.
\ee
Using Eqs.\;(\ref{21}), (\ref{35'})  and (\ref{36}), the evolution  of the  hadron fraction during the phase transition is governed by
\begin{eqnarray}\l{38}
\dot{h}&=&- \frac{(3+n\phi_0\bar{F})(1+rh)}{2r(3\omega+4)\theta\phi_0}
 \Bigg[3\Big( 3\omega-4 \Big)(p_c-\lambda)^2 \phi_0^2 \tilde{f}^2(\phi_0)  \\ &&\;\; + \Big\{ \omega A^2_1+6(2+3\omega+\omega^2)A_2^2
 -6\omega A_2(p_c-\lambda) - 12(p_c - \lambda)^2 \Big\} f^2(\phi_0)\n \\
 &&\;\; + 6\Big\{(4-3\omega)(p_c-\lambda) +2\omega A_2 \Big\}
   (p_c-\lambda)\phi_0 f(\phi_0) \tilde{f}(\phi_0)
  +4(3\omega+4)\phi_0 V(\phi) \Bigg]^{1/2} , \nonumber
\end{eqnarray}
where
$$A_1=3p_c-\rho_Q  -(\rho_H-\rho_Q)h-4\lambda,$$
$$A_2=\rho_Q+(\rho_H-\rho_Q)h+\lambda.$$


\subsection{Evolution of temperature in  the  hadronic era}

In the hadronic phase, the equation of state is given by Eq.\;(\ref{26}), and from
 Eq.\;(\ref{29})  one can obtain
\begin{equation}\l{40}
H=- \frac{3\dot{T}}{(3+n\phi_0\bar{F})T},
\end{equation}
while from the Friedmann equation, (\ref{21}), and Eqs.\;(\ref{26}) and (\ref{40}) we arrive at
\begin{eqnarray}\l{41}
\dot{T}&=&-\frac{ (3+n\phi_0\bar{F})T}{6 (3\omega+4)\theta\phi_0}
   \Bigg[ 3\Big( 3\omega - 4 \Big)(a_{\pi} T^4-\lambda)\phi_0^2 \tilde{f}^2(\phi_0) \\
  &+& 6\Big\{ (4+6\omega+3\omega^2)a^2_{\pi}T^8 \omega(1+\omega)\lambda a_{\pi}T^4 +\omega(6+\omega)\lambda^2 \Big\}f^2(\phi_0)\n\\ &&+ 4(3\omega+4)^2 \phi_0 V(\phi_0)
   +6\Big\{(3 \omega +4)a_{\pi}T^4+\lambda(5\omega -4) \Big\}(a_{\pi}T^4 -\lambda)\phi_0 f(\phi_0) \tilde{f}(\phi_0)
  \Bigg]^{1/ 2}\n.
\end{eqnarray}


\section{An example}
In this Section we  study a model with definite, simple,  functional forms for  $f(\phi)$ and $V(\phi)$. Two  scalar field  potential energy densities, exponential and inverse power law, are commonly  used in discussion of  the chameleon mechanism. Here we consider  the inverse  power law potential energy density \cite{23a}
\begin{equation}\label {42}
V(\Phi)=M^5 \left( \frac{{M^2}}{\phi}\right)^{\xi},
\end{equation}
where $\xi>0$, $M$ is a constant  mass scale and  the scalar field, $\phi$, has ${\rm mass}^2$  dimension. The authors of Ref.\;\cite{7} consider the solar system constraints  for a model  with this potential and find that for small values of $\xi  \in (0 ,  2)$  the magnitude of $M$ is  $\sim 10^{-3}$ eV. Therefore, the potential may be written as
\begin{equation}\l{43}
V(\phi) \sim {10^{-3(2\xi +5)} \over \phi^{\xi}}   \;\ {\rm eV}^5 .
\end{equation}
 Since the characteristic energy density scales of   other quantities such as  $\rho$, $p$, $\lambda$, and  constants of the model,    are of order an  MeV,  the scalar field potential energy density term is very small compared  to other terms in the Lagrangian density and we can  ignore  it. Also   to make the equations tractable, we consider a simple functional form  for $f(\phi)$, $f(\phi) = \phi$.
\subsection{Evolution of   temperature  in the QGP for general $U(T)$}

  To determine   the relevant quantities  we use  equations derived in Sec.\;2. Matching   $f(\phi)=\phi $ and  $ \tilde{f}(\phi) = 1 $, Eq.\;(\ref{29})  becomes
 \begin{equation}\l{44}
\dot{\rho_q}+H(3+n)(\rho_q +p_q)=0,
\end{equation}
and the Hubble parameter is given by

\begin{eqnarray}\l{45}
H= -\frac{ 2(3a_q-\alpha_T)T^2 +\gamma_T  }{2(3+n)a_q T^3}\dot{T},
\end{eqnarray}

Integrating  Eq.\;(\ref{45})   gives the scale factor
\begin{equation}\l{50}
a(T) =C T^{-K} e^{B_0/2T^2},
\end{equation}
where $C$ is a constant of integration and the other constants are
\begin{eqnarray}\l{51}
K&=&-\frac{3a_q -a_T}{(3+n)a_q },\\
B_0 &=&{\gamma_T \over 2(3+n)a_q},\l{51'}
\end{eqnarray}
Also, the Friedmann equation is
\begin{eqnarray}\l{54}
H^2&=& \frac{1}{4(3\omega+4)^2\theta^2} \big(12 +19\omega +6 \omega^2 \big)\rho^2.
\end{eqnarray}

  Combining  Eqs.\;(\ref{45}), (\ref{50}) and (\ref{54}), the equation for  $\dot{T}$  in the quark gluon  phase (QGP) is
\begin{eqnarray}\l{55}
\dot{T}&=&-{(3+n)a_q\over (3\omega +4)}\sqrt{{12 +19\omega +6 \omega^2\over 6+6n-\omega n^2}}\;\;\;\Bigg\{ \frac{ T^3\Big[(3a_q-a_T)T^4 +2\gamma_TT^2 +B +\lambda\Big]}{ 2(3a_q-\alpha_T)T^2 +\gamma_T   }\Bigg\}.
\end{eqnarray}
   We   numerically integrate   this equation   and the results are shown   in Fig.\;1.{  Figure\;1a  shows the decreasing rate  of
 temperature as a function of  cosmic time, $\tau$, in quark-gluon phase (QGP),     for different values of $\omega$, with $n  = 0.015 $, $N_0 = 2 \times 10^5$, and $\lambda = 10\times 10^8 $ MeV$^4$. This plot shows that by increasing $\omega$  the decreasing rate of temperature will be faster and  this decreasing is  occurred at about $(0.05-0.25)$  nanosecond after  the big bang when $T=T_c \approx 125 $ MeV. Figure\;1b shows  the scale factor  as a function of  temperature, $T$,   in QGP and it clearly indicates an expanding Universe at that time.}
\begin{figure}[ht]\label{1}
\begin{minipage}[b]{1\textwidth}
\subfigure[\label{fig1a} ]{ \includegraphics[width=.45\textwidth]%
{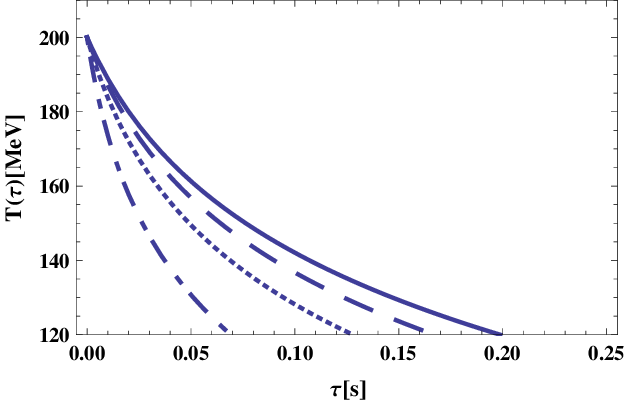}} \hspace{1cm}
\subfigure[\label{fig1b} ]{ \includegraphics[width=.45\textwidth]%
{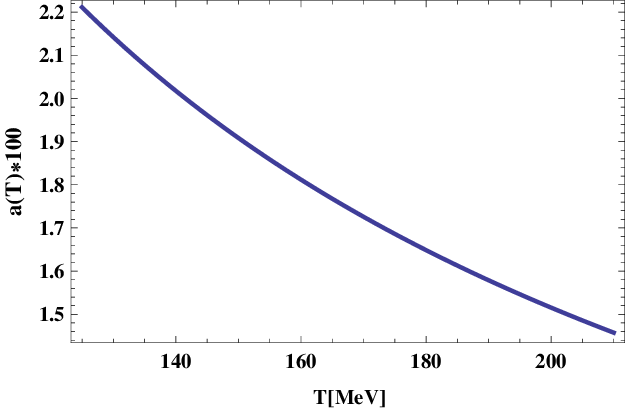}}
\end{minipage}
\caption{ (a) Temperature  as a function of cosmic time in the QGP for  $\omega =10^4$ ({\rm solid line}), $1.5\times 10^4$ ({\rm dashed}), $2\times 10^4$ ({\rm dot}), and $2.5\times 10^4$ ({\rm dot-dashed}). (b) Scale factor  as a function of  temperature in  the QGP for general  $U(T)$. We have set $N =2\times 10^5 $, $\lambda = 10^9 $ Mev$^4$, $n = 0.015$, and  $B^{1/4} = 200  $ Mev.  }
\end{figure}


\subsection{Evolution of temperature in the  QGP for $U (T) = B$}
By   matching $f = \phi $ in  Eq.\;(\ref{34}), we have
\begin{equation}\l{56}
H=- \frac{3}{(3+n)}{\dot{T} \over T}.
\end{equation}
Integrating of this equation gives the scalar field as a function of temperature,
\begin{equation}\l{57}
a(T) =c T^{-3/(3+n)}.
\end{equation}
Eq.\; (\ref{32}) is became
\begin{eqnarray}\l{57'}
H^2&=& \Bigg( \frac{12+19\omega+6\omega^2}{6+6n-\omega n^2} \Bigg) \frac{(\rho_q+\lambda)^2}{4(3\omega+4)^2}
\end{eqnarray}
Using Eqs.\;(\ref{24}),  (\ref{25}),(\ref{56}), and   (\ref{57'})  one  obtains an expression for   $\dot{T}$
\begin{eqnarray}\l{58}
\dot{T}&=&-\frac{(3+n)}{6(3\omega+4)} \sqrt{\frac{12+19\omega+6\omega^2}{6+6n-\omega n^2}} \Big[T\big( 3a_qT^4+B+\lambda\big) \Big]
\end{eqnarray}

 We  numerically solved this equation   and  plot the result    in Fig.\;2. {  Figure\;2a  shows the decreasing   of
 temperature as a function of  cosmic time, $\tau$, in quark-gluon phase (QGP),     for different values of $\omega$, with $n  = 0.015 $, $N_0 = 2 \times 10^5$,  $\lambda = 10\times 10^8 $ MeV$^4$ , and $U(T)=B$. This plot shows that by increasing $\omega$  the decreasing  of temperature will be faster and  this decreasing is  occurred at about $(0.03-0.08)$  nanosecond after  the big bang when $T=T_c \approx 125 $ MeV. Figure 2a indicates in the $U(T)=B$ case the QGP is occurred earlier than general case, $U(T)= B+\gamma_T T^2 - \alpha_T T^4$.   Figure\;2b shows  the scale factor  as a function of  temperature, $T$,   in QGP  in the $U(T)=B$ case  and it clearly indicates an expanding Universe at that time.}
\begin{figure}[ht]\label{2}
\begin{minipage}[b]{1\textwidth}
\subfigure[\label{fig1a} ]{ \includegraphics[width=.45\textwidth]%
{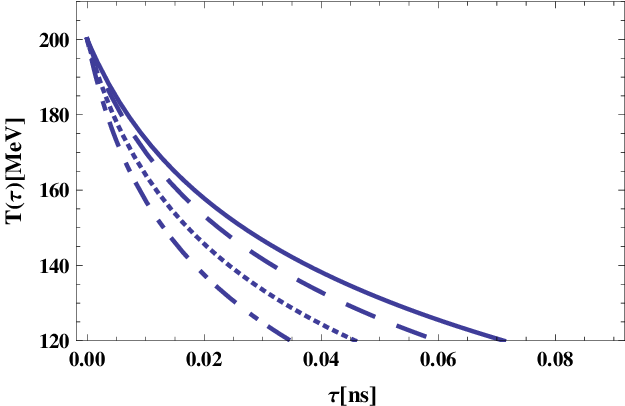}} \hspace{1cm}
\subfigure[\label{fig1b} ]{ \includegraphics[width=.45\textwidth]%
{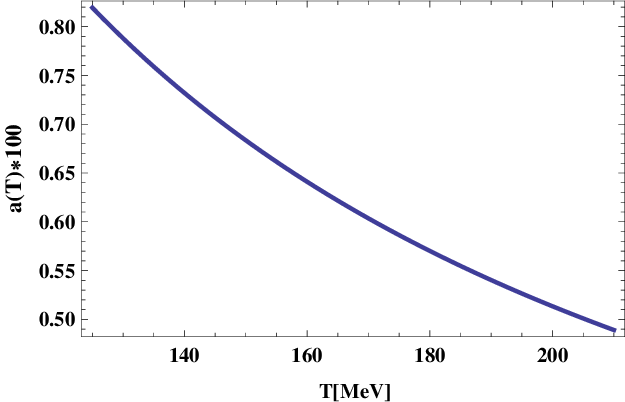}}
\end{minipage}
\caption{ (a) Temperature  as a function of cosmic time in the  QGP for $\omega =8\times10^3$ ({\rm solid line}), $1.4\times 10^4$ ({\rm dashed}), $2\times 10^4$ ({\rm dotted}), and  $2.6\times 10^4$ ({\rm dotted-dashed})  and for $U(T)= B$. (b) Scale factor  as a function of  temperature  in QGP. We have set $N =2\times 10^5 $, $\lambda = 10^9 $ Mev$^4$, $n = 0.015$, and  $ U(T) = B^{1/4} = 200  $ Mev.    }
\end{figure}


\subsection{Evolution of hadron volume fraction}
 As  mentioned above the pressure  during the phase transition is constant, $p_c \approx 4.6 \times 10^8$MeV$^4$ and the density of quark matter and hadron matter at the transition  are  $\rho_q \approx 5 \times 10^9$MeV$^4$, $\rho_h\approx 1.38 \times 10^9$ MeV$^4$, respectively.
Therefore, by matching  $f(\phi)=1 $ in   Eq.\;(\ref{36}) we have
\begin{equation}\l{60'}
H=- \frac{r\dot{h}}{(3+n)\big(1+rh\big)}
\end{equation}
where
\bea\l{59}
r&=&\frac{\rho_H-\rho_Q}{p_c+\rho_Q} = -5.6\times 10^{-6},
\eea

Integrating Eq.\;(\ref{60'})  gives   the scale factor on the brane as a function of
the hadronic volume  fraction, $h(\tau)$,
\begin{equation}\l{60}
a(\tau) =a(\tau_c) {\biggr [}1+rh(\tau) {\biggl ]}^{-{1 /(3+n)}},
\end{equation}
here we have  assumed $h(\tau_c) =0$. So, using Eq.\;(\ref{54}), the time evolution  equation of  the matter fraction in the hadronic phase is
\begin{eqnarray}\l{61}
\dot{h}&=&-\frac{(3+n)}{2(3\omega+4)} \sqrt{\frac{12+19\omega+6\omega^2}{6+6n-\omega n^2}} \ \frac{(1+rh)}{r} \Big[ (1+m h)\rho_Q + \lambda \Big]
\end{eqnarray}
 Numerically evaluated  $h(\tau)$'s are presented in Fig.\;3 for various  values of $\omega$. {  Figure\;3a shows  the hadron volume fraction   during the QHPT for  $\omega =10^4$ ({\rm solid line}), $1.5\times 10^4$ ({\rm dashed}), $2\times 10^4$ ({\rm dotted}), $2.5\times 10^4$ ({\rm dotted-dashed}),  as a function of  cosmic time. This figure indicates that by increasing the dimensionless parameter of BD model, $\omega$, the rate of quark-hadron phase transition  will be faster and QH phase transition  takes about ($0.1 -0.5$) nanosecond in a constant temperature.   Figure\;3b shows the scale factor of the universe  during the QHPT  as a function of the  hadron volume  fraction. It is well known that when the QHPT  accurse  the density of quark gluon plasma  decreases but the  hadron volume fraction   and the scale factor of universe  increase. Moreover   Fig.\;3b states  that  during the QH phase transition the Universe is expanding. }
\begin{figure}[ht]\label{3}
\begin{minipage}[b]{1\textwidth}
\subfigure[\label{fig1a} ]{ \includegraphics[width=.45\textwidth]%
{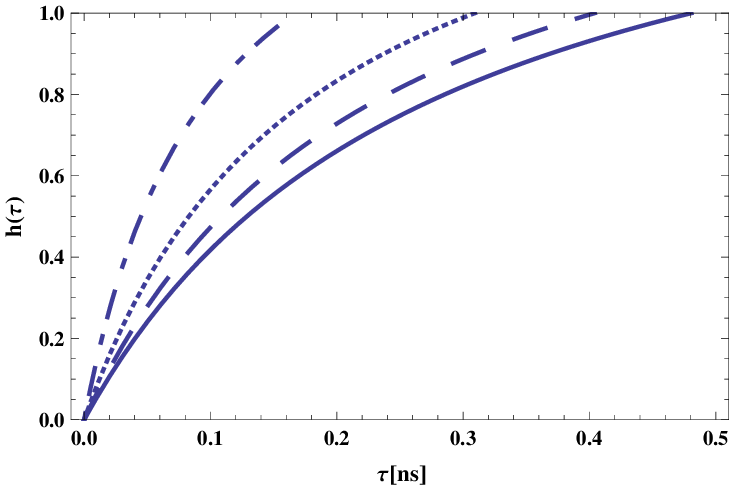}} \hspace{1cm}
\subfigure[\label{fig1b} ]{ \includegraphics[width=.45\textwidth]%
{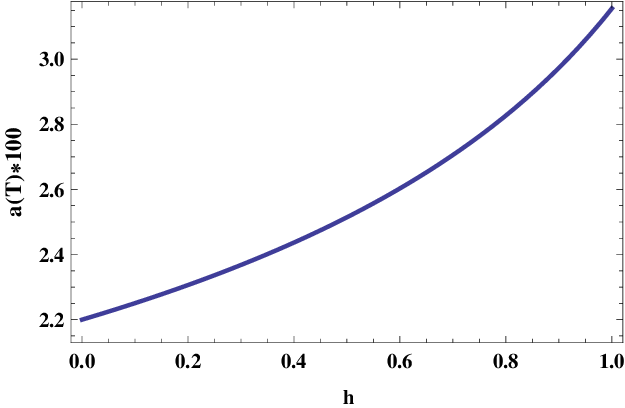}}
\end{minipage}
\caption{(a) Hadron volume fraction as a function of cosmic time for  $\omega =10^4$ ({\rm solid}), $1.5\times 10^4$ ({\rm dashed}), $2\times 10^4$ ({\rm dotted}), and $2.5\times 10^4$ ({\rm dotted-dashed}). (b) Scale factor  as a function of hadron  volume fraction. We have set $N =2\times 10^4 $, $\lambda = 10^9 $ Mev$^4$, $n = 0.015$, and  $  B^{1/4} = 200  $ Mev.   }
\end{figure}


\subsection{Evolution  of temperature in  the hadronic area}
Using the equation of state, Eq.\;(\ref{26}), and the energy  conservation relation, Eq.\;(\ref{29}), we have
\begin{equation}\l{63}
H=- \frac{3\dot{T}}{(3+n)T}.
\end{equation}
Integrating  this equation gives
\begin{equation}\l{64}
a(T) = cT^{-3 / (3+n)} ,
\end{equation}
and from the Friedmann equation one arrives at
\begin{eqnarray}\l{65}
\dot{T}&=&-\frac{(3+n)}{6(3\omega+4)} \sqrt{\frac{12+19\omega+6\omega^2}{6+6n-\omega n^2}} \ \Big[ T(3a_{\pi}T^4+\lambda) \Big]
\end{eqnarray}
  We numerically solved  Eq.\;(\ref{65}) and  plot the results  in Fig.\;4. {  Figure\;4a shows  the temperature as a function of cosmic time in the hadron phase for  $\omega =10^4$ ({\rm solid line}), $1.5\times 10^4$ ({\rm dashed}), $2\times 10^4$ ({\rm dotted}), $2.5\times 10^4$ ({\rm dotted-dashed}). This figure indicates that by increasing the dimensionless parameter of BD model, $\omega$, the rate of decreasing of temperature  will be faster and the hadron phase is occurred about $(1.2-2.5) ns $    after the big bang. This result is in a good agreement with the expanding Universe in Fig.\;1, general case of $U(T)$.   Figure\;3b shows the scale factor of the universe  in hadron phase   as a function of cosmic time and indicates an expanding Universe in this phase.  }
\begin{figure}[ht]\label{4}
\begin{minipage}[b]{1\textwidth}
\subfigure[\label{fig1a} ]{ \includegraphics[width=.45\textwidth]%
{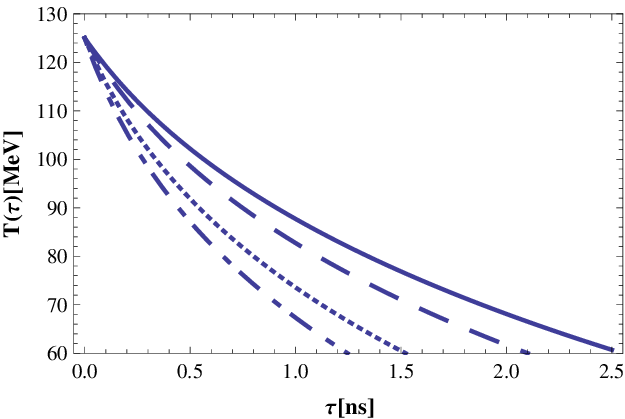}} \hspace{1cm}
\subfigure[\label{fig1b} ]{ \includegraphics[width=.45\textwidth]%
{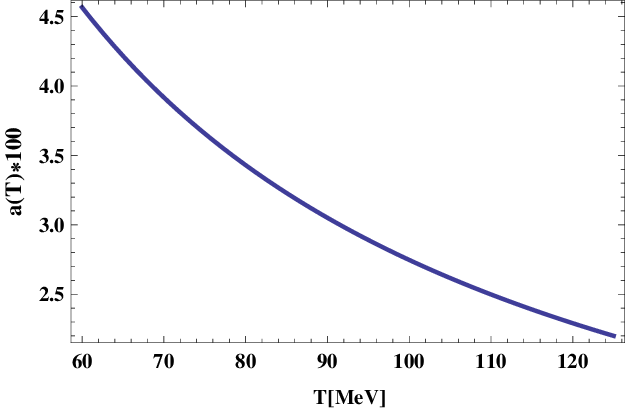}}
\end{minipage}
\caption{(a) Temperature as a function  of cosmic time  in the  hadronic phase  for  $\omega =10^4$ ({\rm solid line}), $1.55\times 10^4$ ({\rm dashed}), $2.1\times 10^4$ ({\rm dotted}),  and $2.55\times 10^3 $({\rm dotted-dashed}). (b) Scale factor  as a function of  temperature in the  hadronic phase. We have set $N =2\times 10^5 $, $\lambda = 10^9 $ Mev$^4$, $n = 0.015$, and  $  B^{1/4} = 200  $ Mev.   }
\end{figure}

 The effective temperature is plotted in Fig.\;5 for various  $(\omega, n)$ for which $\omega n = c$. Figures 5a and 5b  show $T(\tau)$ for $(\omega , n) = (10^5, 10^{-3})$, and  $(3\times10^4,10^{-2}/3)$  with  $\omega n = 100$, and $(\omega , n) = (10^6,10^{-3})$, and $(3\times10^5,10^{-2}/3)$  with $\omega n = 1000$  respectively. It is seen that these curves are slightly different.  We found  that for $n \leq 0.015 $, $c\leqslant 120$,  and $100\leqslant \omega \leqslant 10^6$, all curves are very similar function of    cosmic time while  for $c > 120 $ the temperature curves are differ slightly.
\begin{figure}[ht]\label{5}
\begin{minipage}[b]{1\textwidth}
\subfigure[\label{fig1a} ]{ \includegraphics[width=.45\textwidth]%
{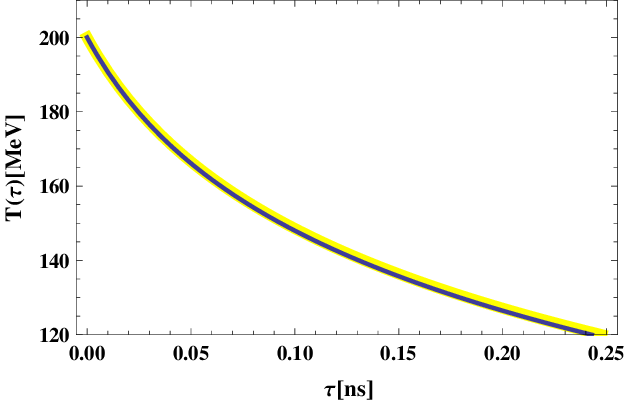}} \hspace{1cm}
\subfigure[\label{fig1b} ]{ \includegraphics[width=.45\textwidth]%
{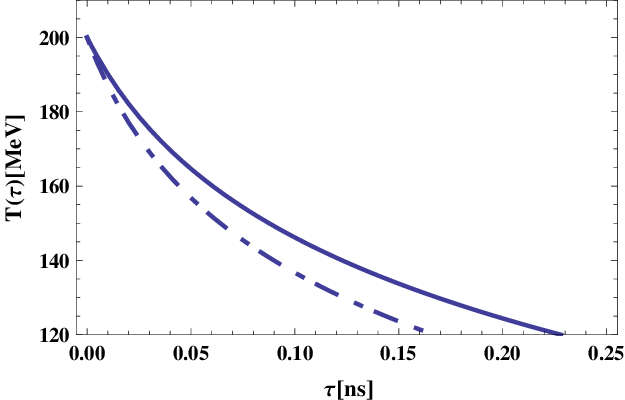}}
\end{minipage}
\caption{(a) Temperature as a function of cosmic time  in  the QGP for  $(\omega, n) = (10^5, 10^{-3})$, and  $(3\times10^4,10^{-2}/3 )$ for  which $\omega \times n = 100.$ (b) Temperature in the QGP for   $(\omega, n)=(10^6,10^{-3})${\rm solid line}), and $3\times10^5, 0.01/3$({\rm dashed line}) for which $\omega \times n = 1000$. We have set $N =2\times 10^5 $, $\lambda = 10^9 $ Mev$^4$, and  $B^{1/4} = 200 $ Mev.   }
\end{figure}

\section{QCD phase transition }
  Depending  on the values of the quark masses, the phase transition in QCD,  characterized by the
singular behavior of the partition function, could  be a first or second order phase transition, or it could  be only a crossover with rapid changes in some observables. In this Section we  examine  physical quantities related to the  quark-hadron phase transition, based on the assumption of a  smooth crossover approach,  in the  CBD model of the  brane world scenario.

As   mentioned earlier, to  study the quark-hadron phase transition we need  the equation of state of matter  in both the  quark and the  hadron phase regimes. Different approaches have been used to obtain the equation of state. Recently,  detailed computations  of the equation of state  have been
performed using the  fermion formulation on lattices with temporal extent $N_t = 4, 6$ \cite{7a, 8a},
 $N_t=8$ \cite{9a} and $N_t= 6, 8, 10$ \cite{10a}. In the high temperature region, where $T > 250$ MeV, the trace anomaly can be precisely calculated. So one can use the lattice data for the trace anomaly in the high temperature  to construct a realistic equation of state. On the other hand  in the low temperature region, where $T \lesssim 180$ MeV,   the trace anomaly is affected by  large discretization effects, but the  hadronic resonance gas (HRG) model can be used to determine a  realistic low temperature equation of state  \cite{36s}.
\subsection{High temperature region}
As mentioned above,  lattice data for the trace anomaly can be used to determined the  equation of state at high temperature, $T > 250$
MeV \cite{36s}. In this regime  the gluons and quarks are effectively massless so  behave like   radiation,   and one can fit the lattice data to a simple equation of state
\bea\l{66}
\R(T)&\thickapprox & \alpha_{r}T^4,\\
p(T) &\thickapprox & \sigma_{r}T^4.\l{67}
\eea
Here $\alpha_{r} = 14.9702\pm 009997$ and $\sigma_{r} = 4.99115\pm 004474$ are found from a least squares fit \cite{8a}.
Substituting Eqs.\;(\ref{66}) and (\ref{67}) into Eq.\;(\ref{29}), we obtain
\begin{equation}\l{68}
H=-{ 4 \alpha_r \dot{T}\ov (3+n)(\alpha_r+\sigma_r)T }.
\end{equation}
Integrating  Eq.\;(\ref{68}) we have
\begin{equation}\l{69}
a(T)= a_c T^{-{ 4 \alpha_r/ (3+n)(\alpha_r+\sigma_r) }},
\end{equation}
where $a_c$  is the constant of integration and,  by using Eqs.\;(\ref{32}), (\ref{66}), (\ref{67}), and (\ref{68}),  $\dot{T}$ is
\begin{eqnarray}\l{70}
\dot{T}&=&-\frac{(3+n)(\alpha_r + \sigma_r)}{8\alpha_r (3\omega+4)} \sqrt{\frac{12+19\omega+6\omega^2}{6+6n-\omega n^2}}\;  T(\alpha_r T^4 + \lambda) .
\end{eqnarray}

\begin{figure}[ht]\label{0}
\centerline{ \includegraphics[width=7cm] {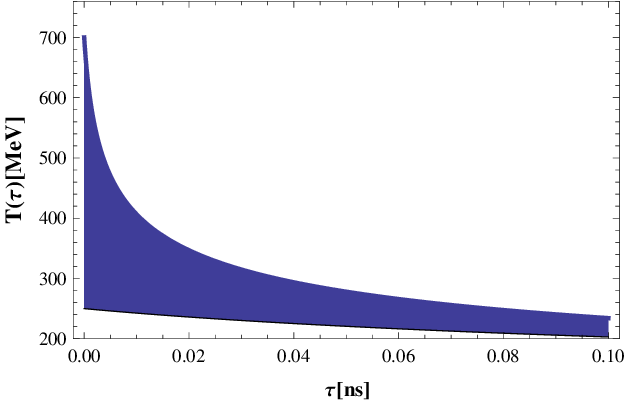}}
\caption{   $T$ as a function of cosmic time,  $\tau$, for   $250 < T < 750$ MeV.  We have set $N =2\times 10^5 $ and  $\lambda = 10^9 $. }
\end{figure}

We  numerically integrate  Eq.\;(\ref{70}) and  plot the results in Fig.\;6 for two initial values of $T$, $T_{01}=250$ MeV and $T_{02}=700$ MeV.  This figure shows  the effective  temperature in the QGP  in the  CBD model of brane gravity for $T >250$ MeV,   obtained for the  smooth crossover approach. We see that  the Universe  become cooler and the temperature drops to 250 MeV   at about $0.02-0.1 $ ns after the big bang. Since in this regime  for $T_{01}\sim250$ the temperature is almost independent  of cosmic time,  matter remains in quark-gluon phase above  $T\sim 250$ MeV.

\subsection{Low temperature region}

As mentioned above, the hadronic resonance gas (HRG) model can be used  to build a realistic equation of state at low temperatures, $T \lesssim 180$ MeV \cite{36s}.  In the  HRG scenario QCD  is treated
as a non-interacting gas of fermions and bosons \cite{37s}. In fact, the fermions and bosons in this model are
 mesons and baryons. The basic idea of the HRG model is to
implicitly account for the strong interaction in the confinement phase by looking only  at  hadronic
resonances, since these are  the relevant low temperature  degrees of freedom.
In this regime, it is believed  that the  HRG model  provides  a reasonable  description of thermodynamic quantities.

 The HRG result can also be parameterized for the trace anomaly  as \cite{36s}
\be\l{71}
{I(T) \ov T^4}=  {\R -3p \ov T^4} = a_1T + a_2T^3 + a_3T^4 + a_4T^{10},
\ee
where $I(T)=\R(T) -3p(T)$ is the trace anomaly,  $a_1$ = 4.654 GeV$^{-1}$, $a_2 = - 879$ GeV$^{-3}$, $a_3$ = 8081 GeV$^{-4}$, $a_4 = - 7039000 $ GeV$^{-10}$.
In lattice QCD, through the computation of the trace anomaly $I(T) $, one can estimate the pressure, energy density, and entropy density,  with the help of the thermodynamics identities. The pressure difference at two temperatures $T$ and $T_{\rm low}$ is an integral of the trace anomaly
\be\l{72}
{p(T) \ov T^4 } - {p(T_{\rm low})\ov T^4_{\rm low}} = \int^T_{T_{\rm low}} {dT' \ov T'^5}I(T').
\ee
By choosing a sufficiently small lower integration limit, $p(T_{\rm low})$ can be neglected due to the exponential
suppression. The energy density $\R(T) = I(T) + 3p(T)$  can be computed. This procedure is known as the integral method \cite{39s}.

 Using
Eqs.\;(\ref{71}) and (\ref{72}) we obtain
\be\l{73}
\R(T)= 3a_0T^4 + 4a_1T^5 + 2a_2T^7 + {7a_3 \ov 4}T^8 + {13a_4 \ov 10} T^{14},
\ee
\be\l{74}
p(T) =a_0T^4 + a_1T^5 + {a_2 \ov 3}T^7 + {a_3 \ov 4}T^8 + {a_4 \ov 10} T^{14},
\ee
where $a_0= -0.112$. In this step, we consider matter before phase transition at low temperatures when quarks become confine   as non-interacting gases of fermions and bosons \cite{37s}. From the conservation relation during this epoch,  we have
\begin{equation}\l{75}
H=- \frac{ \big [12a_0 T^3+20a_1 T^4 + B_0(T)\big]\dot{T}}{(3+n)\big[4a_0 T^4 + 5 a_1 T^5 + B_1(T)\big]},
\end{equation}
where
\begin{eqnarray}
B_0(T)&=& 14a_2 T^6 + 14 a_3T^7 + \frac{92}{5} a_4 T^{13},\l{76} \\
B_1(T)&=& \frac{7}{3}a_2T^7 + 2a_3T^8 + \frac{7}{5}a_4T^{14}.\l{77}
\end{eqnarray}

To obtain the  scale factor as a function of temperature we must integrate Eq.\;(\ref{75}).
This can be re-expended the time derivative of temperature
\begin{eqnarray}\l{79}
\dot{T}&=&-\frac{(3+n)}{2(3\omega+4)}  \sqrt{\frac{12+19\omega+6\omega^2}{6+6n-\omega n^2}} \\
 & &  \times \Bigg\{ \frac{4a_0 T^4 + 5 a_1 T^5 + B_1(T)}{12a_0 T^3+20a_1 T^4 + B_0(T)}
 \Big( 3a_0T^4 + 4a_1T^5 + 2a_2T^7 + {7a_3 \ov 4}T^8 + {13a_4 \ov 10} T^{14} + \lambda \Big) \Bigg\} \nonumber
\end{eqnarray}
 We     numerically  integrate Eq.\;(\ref{79}) and plot the result   in Fig.\;7  for two initial values $T_{01}=50$ MeV and $T_{02}=180$ MeV. Figure\;7  shows
 temperature as a function of the cosmic time, $\tau$, in the low energy region for the  CBD model of brane gravity. This figure shows that, in the low temperature regime of the QCD phase transition ( for a crossover transition where HRD is used),  the QGP  of the Universe is about 1-10  nanoseconds after the big bang. It is seen that in this regime  for $T_{01}\sim30$ MeV the temperature is almost independent of cosmic time and therefore    the area which the QGP is occurred in in the interval MeV $30 \lesssim T \lesssim 180$ MeV. One can clearly seen that the QGP in the low region of the smooth crossover approach is occurred later than first order phase transition formalism.
 \begin{figure}[ht]\label{0}
\centerline{ \includegraphics[width=7cm] {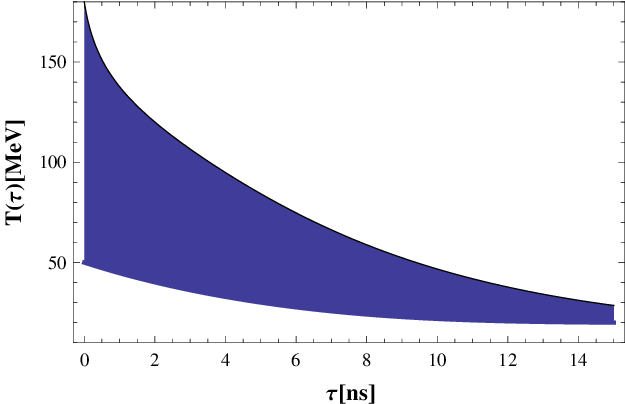}}
\caption{  $T$ as a function of  cosmic time, $\tau$,  in the interval   $50 < T < 180 $ MeV in the CBD model of brane gravity. We have set $\omega = 10^4 $, $ n= 0.002$,  and  $\lambda = 10^9$MeV$^4$.   }
\end{figure}

\section{Conclusion}
In this paper we have studied the quark-hadron phase transition in a chameleon Brans-Dicke   brane-world scenario.
 We have investigated  the evolution of  quantities relevant to the physical
description at early times, such as the  energy density, temperature and scale factor, before, during, and
after the phase transition.
We have found that for  $n < 0.015 $ and $100\leqslant\omega \leqslant10^6$ phase transition occur and  as increases times
the  effective temperature of the quark-gluon plasma and  the hadronic fluid   decrease. We have  ploted  the effective temperature and  scale factor of the Universe at  different stages of the phase transition for various values of $\omega$ and $n$. All  plots show that the effective temperature and the FLRW  scale factor  decrease and increase, respectively as  time passes. { Of especial interest is   the increasing  behavior of the scale factor during the phase transition in first order formalism  which indicates that at this stage the Universe is expanding although the  temperature and pressure of the Universe is constant. Our analysis  in the first order phase transition formalism shows that the QGP  takes  place at about $0.05-0.2$ ns  after the big bang and phase transition takes about $0.1-0.5$ ns  and after that we have the hadronic phase at about $1.2-2.5$ ns  after the big bang. }

We compared our results with the results presented in \cite{15, 16, 17}. In \cite{15} the authors studied quark–
hadron phase transition in a Randall – Sundrum  brane model and have shown that for different values of
the barne tension,  $\lambda$, phase transition occurs about at $10^{-6}$s after the big bang. Also in \cite{16, 17}, the authors investigated the quark-hadron
phase transition in a brane-world scenario where the localization of matter on the brane is achieved
through the action of a confining potential and have shown that for different values of  parameters in
their model, phase transition takes place. They found that for various value of $\omega$ the phase transition  has taken  place about microsecond after the big bang, but our investigation shows that the quark-hadron phase transition has occurred  at about  nanosecond  after the  big bang. This is a difference between the results of our study and the studies of other researchers. { This means that due to the interaction between scalar field and matter has made  a brane-bulk energy transfer and   conservation equation of energy density has been  modified. Actually this phenomena change the functionality of the effective  temperature with respect to  cosmic time and  therefore   the rate of expansion of the Universe is increased at the early times.}

At last, we studied the smooth crossover approach for quark-hadron phase transition in  high and low region of temperature in Sec.\;5. We have used the equation of state which are obtained from lattice QCD data. The results of our calculations show that the general behavior of temperature in both of approaches ( smooth crossover and first order phase transition)  is  similar,  although the differences in the energy should be taken into account. { In fact, by considering in detailed, one can see that  the dropping of temperature in the QGP phase of the Universe in the  first order phase  transition approach is slower than the high temperature  region  of the smooth crossover formalism which lattice QCD is used to investigate  the equation of state and faster than the low temperature regime of QCD phase transition (crossover transition) where HRD is used for obtaining the matter equation of state.}
\section{Acknowledgement}
 The work of  Kh. Saaidi has been supported financially  by the University of Kurdistan, Sanandaj, Iran, and  he would like thank to the University of Kurdistan for supporting him in his sabbatical period. The work of B. Ratra was supported  by DOE grant DEFG030-99EP41093 and NSF grant AST-1109275.


\end{document}